# Toward Conceptual Modeling for Propositional Logic: Propositions as Events

**Sabah Al-Fedaghi***

salfedaghi@yahoo.com, sabah.alfedaghi@ku.edu.kw

*Abstract* – Applying logic in the area of conceptual modeling has been investigated widely, yet there has been limited uptake of logic-based conceptual modeling in industry. According to some researchers, another formalization of such tools as EER or UML class diagrams in logic may only marginally contribute to the body of knowledge. This paper reflects on applying propositional logic language to a high-level diagrammatic representation called the thinging machines (TM) model. We explore the relationship between conceptual modeling and logic, including such issues as: What logical constructs model? How does truth fit into the picture produced in conceptual modeling as a representation of some piece of the world it is about? The ultimate research objective is a quest for a thorough semantic alignment of TM modeling and propositional logic into a single structure. Examples that involve the application of propositional logic in certain areas of reality are TM remodeled, where propositions are viewed as TM regions or events. As it turned out, TM seems to shed light on the semantics of propositions. In such a conceptual framework, logical truth is a matter of how things are in actuality and how falsehood is in subsistence. The results show that propositional logic enriches the rigorousness of conceptual descriptions and that the TM semantic apparatus complements propositional logic by providing a background to the given set of propositions. Semantics matters are applied to propositional constructs such as negative propositions, disjunctions, and conjunctions with negative terms.

## I. INTRODUCTION

According to [1],

Systems' complexity has intensified, leading to a rise in developing models using different formalisms and diverse representations even within a single domain. Conceptual models have become larger; languages tend to acquire more features, and using different modeling languages for different components is not unusual. This diversity has caused problems with consistency between models and incompatibility with designed systems.

From this perspective, conceptual modeling needs to adopt a broader perspective to be capable of supporting a wider

-----------------------
**\***Computer Engineering Department, Kuwait University, Kuwait. Retired June 2021, seconded fall semester 2021/2022

range of applications such that "modeling solutions should be usable by as many people and design agents as possible and for as many purposes as possible" [2]. According to [2], an opportunity exists for a much deeper integration of conceptual modeling with other topics by positioning it as the unique language for such an application.

Conceptual modeling involves capturing, abstracting, and representing relevant aspects of reality [3]. 'Reality,' here, refers to the sensible world surrounding us. The common core of how humans represent reality makes it possible to conceptually model diverse, complex, and emerging systems and domains for various people [2]. Consequently, conceptual modeling is necessary to facilitate modeling any domain and can be used by the broadest audiences possible with minimal or no training requirements [2]. In this paper, conceptual modeling is viewed as a holistic, unifying scheme representing real-world systems using a single language called thing machines (TM).

### A. TM in Glance

TM is a scheme for diagrammatically representing real-world systems in an attempt to develop a "universal core" [2] of multidisciplinary ontological-based modeling methodology of such reality phenomena as actuality, existence, substantiality, etc. [4]. Reality is regarded as all that exists and subsists and is modeled in terms of five actions: create, process, release, transfer, and receive. The structure of reality is viewed in two levels: potentiality/actuality organization. Actuality is where potential things are realized, and potentiality is where unactualized things are relegated.

These two levels are associated with the mode (existence/subsistence) of entities called thimacs (Thing/Machine). Entities of actuality are *event* thimacs, and corresponding entities of potentiality are *region* thimacs. Events exist, and regions subsist. A region of subsistence becomes an event when the region is juxtaposed with time.

### B. Conceptual Modelling and Logic: Needs

One of the features of conceptual modeling languages is providing notations that support inferences and formal modeling techniques, especially in emerging applications such as artificial intelligence and social media. Logic can provide the theoretical foundations of such features in conceptual modeling. In general, incorporating logic in conceptual modeling brings precision to the representation since logic gives the graphical elements formal semantics and automated reasoning [5]. Note that logic is sometimes distinguished from reasoning [6]. Additionally, conceptual modeling is a crucial software development that includes requirements analysis, specification, design, implementation, verification and validation, evaluation, and maintenance. Most of these phases may benefit from using formal methods [7].

### C. Logic Applied in Conceptual Modeling

A good review of the topic *logic in conceptual modeling* can be found in [5]. Conceptual modeling has been investigated widely and is crowded with many attempts to assign semantics, yet there has been limited uptake of logic-based conceptual modeling in the industry. According to [5], "another formalization of plain EER, ORM2, or UML class diagrams in yet another logic may only make a marginal contribution to the body of knowledge." Logic as a unifier for conceptual modeling has been explored, albeit with limited tools and mainly covering a small set of constraints [5]. Conceptual data models are also used in Artificial Intelligence to drive the design of intelligent information systems, provided they are given a logic-based specification focusing on such matters as formalizing graphical elements and diagram grammar, examining their features, and whether one could have one logic that maps to all significant diagram-based conceptual modeling languages [5].

Typically, a purely formal approach leaves the issue of its ontological consequence unaddressed. For example, propositional logic does not possess sufficient expressiveness to represent real-world knowledge [8]. According to [2], advances in the 20th century offer valuable tools for theorizing about propositions. The dissatisfaction with this tradition on propositions is that it does not illuminate what a proposition is. According to some sources [2], "Nothing explains why or how propositions are truth-value bearers or representations. The only thing that can be said of propositions is that they are abstract, mind and language-independent entities".

### D. Aims of this paper

In this paper, we are on a mission to explore the relationship between conceptual modeling and logic, including such issues as whether logic tells us anything about the world we try to model. What logical constructs themselves model? Does reality have some logical structure to it? Does logic provide, in some sense, a picture of the world? How does truth fit into the image produced in conceptual modeling as a representation of some piece of the world it is about? The ultimate objective is a search for a thorough semantic alignment of TM modeling and propositional logic into a single structure (See Fig. 1). In Fig 1, the semantic base is an intermediate interpretation of TM and propositional logic notions. The approach in this venture is that instead of a proposition that can be semantically assessed as true or false, we aim to convert it into a diagrammatic representation based on the ontological foundation of TM's two modes reality: actuality, and potentiality. The intended benefit is strengthening the TM model with a touch of formality (rigidness). As it turned out, as a result of this research, TM seems to shed light on the semantics of the proposition. We avoid generalities such as discussion of overarching strategies and broad principles, and present a specific study that involves,

(a) Focusing on propositional logic, the simplest form of logic that assumes the world contains facts.

(b) Application to a particular high-level conceptual modeling language, TM.

TM has been proposed in software engineering and theoretically applied by the author in modeling many applications areas such as networks, hardware, systems, story-telling, communication, railcar systems, robotics, business, etc. (See, for example [9-12]). Focusing on specific logic and a particular proposed conceptual modeling language is a constructive scheme to uncover difficulties in semantics and metaphysical perspectives since the scheme implicates reality. TM modeling can be utilized as a research apparatus for mapping various notations that involve modeling aspects of relevant reality.

In the context of propositional logic, this research includes such tasks in propositional logic, conceptually modeling propositions and logical laws, e.g., the law of contradiction, negativity, logical constants, etc. For example, in logic, the word *model* (in the logical sense) represents a possible state of affairs in the world. In propositional logic, this assignment specifies a truth value (true or false) for each propositional symbol. How such a logical model is related to a conceptual model?

Of course, there have been works to supplement conceptual modeling languages with logical constructs, e.g., enriching UML-based conceptual modeling with formalized constraints using tools of Boolean logic. In contrast, research in this paper attempts a thorough semantic alignment of TM modeling and propositional logic into an intermediate structure.

### E. Look at Logic and Reality

In the last half-century, logic has found radically new and important roles in computation and information processing [13]. According to [14], the relationship between logic and reality has often been complicated. From a particular perspective, logic is listed among prior disciplines; hence, it is independent of reality.

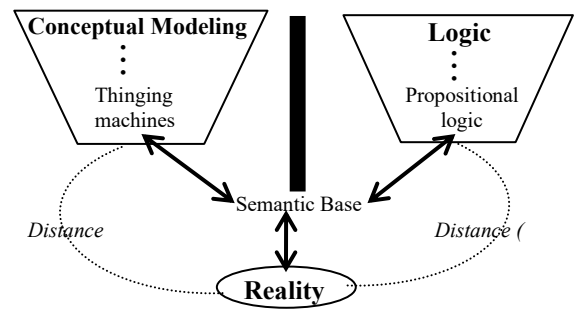

Fig 1. Alignment of TM modeling with propositional logic

Some logicians think that logic is not supposed to go into the metaphysical problems and is just supposed to provide some validity-checking machinery "just like elementary math texts, as such, need not worry about the ontological status of "mathematical entities" when they concern themselves only with providing reliable methods of calculation or construction" [15]. From Lupasco's [16] standpoint (to be discussed later), "all standard logics were bivalent [every proposition has exactly one truth value] and hence incapable of describing or modeling complex energetic processes."

Conversely, in the last half-century, logic has found radically new and important roles in such application subjects as computation and information processing [13]. For example, Knowledge representation systems aim to research techniques of knowledge description and structuring, which should facilitate reasoning tasks. Reasoning is also used in the so-called conceptual schemas (e.g., UML/OCL) in the specification stage of software development. It aims to answer questions regarding what kind of instances a conceptual schema admits [17].

### F.  Content of Paper

This paper unfolds as follows: the next section reviews the TM model in several previous papers (e.g., [9-12]). The section also includes new materials since we further develop the model in each new research. Section 3 gives a TM modeling example with propositional logic (From [18]). The example involves a restaurant where the waiter serves plates according to questions to conclude. Answers received allow the waiter to infer where dishes must go. This inference schema can be represented as (using F for "fish," M for "meat," and V for "vegetarian") F or V or M, not M, not F ⇒ V [18]

Section 4 explores the connections between TM and logical constructs, e.g., propositions, negative propositions, and types of negativity. Section 5 focuses on logical constants (e.g., ∧ and ∨) that relate two TM thimacs to make a new thimac from them. Section 6 includes a case study of TM modeling a library example from [19]. The requirements for borrowing and returning a specific book from a school library are formalized in propositional logic. The book can be in any one of the following four states: on_stack (S), on_reserve (R), on_loan (L), and requested (Q). Constraints s are specified as, S ⇒ ¬(R ∨ L), R ⇒ ¬(S ∨ L), L ⇒ ¬(S ∨ R) and Q ⇒ S ∨ R.

## II.    TM Modeling

TM modeling describes reality as all that *exists* (actual world) and *subsists* (potential world). The potentiality/actuality (Lupasco's terminology—see [16]) divide reflects a contingent being where a being does not exist but still has the potential to exist. In a specific context, it may be elevated from potentiality to actuality while still its potentiality conditions support its actuality. In this situation, potential regions are necessary to understand actuality.

The structure of *reality* is viewed into two levels, with a potentiality/actuality scheme adopting an idea that goes back to the Stoic modes of reality (see Fig. 2—terms in the figure will be defined later). A TM *model* (e.g., a model of a school library) is a particular specification of actual and potential reality. The two TM levels roughly emulate the 'meanings' of simple propositions in propositional logic: truth and falsehood.

### A.  Some Related Views of Reality as Actuality and Potentiality

The potentiality/actuality divide is rooted in Aristotle's matter, which refers to the fundamental substance that provides the potentiality for a thing to become actualized in a specific form. The actual being is a being in total, and the potential being lacks fullness [20]. According to Aristotle, both the actual and the potential being are due to every entity, which means that reality and virtuality [roughly, TM potentiality] are distributed among the beings of our world instead of their concentration into entirely separate worlds [20]. In TM, subsistence and existence are superimposed like a double-image impression (e.g., Rubin's vase) [21].

Similar ideas may be traced over history to many works in ontology. For example, Plato advocated the physical world as a reflection of the world of forms with one form (a form of the good) out of which everything manifests. In modern times, Wittgenstein distinguished the world as "the totality of states of affairs that obtain" from actuality, which, he tells us, is "the obtaining and non-obtaining of states of affairs" (see [22])."

The TM two-level model finds its roots in Lupasco's 'onto-logics' where (almost) everything is logical, and thus logic is to be found in everything [23]. Lupasco's conception of logic is that reality has a rational form, which implies a basis for logic in real-world phenomena and its use for their description [16]. According to [16], "Lupasco's basic insight: logic not only should but can be extended to reality, provided one takes into account, and gives proper metaphysical weight to, some of its characteristics that have tended to be neglected." For Lupasco, passage from potentiality to actuality and passage from actuality to potentiality are acting on each other [16]. Lupasco provided a rejection of contradiction where every event was always associated with a 'no event,' such that the actualization of an event entails the potentialization of a 'no event' and vice versa without either ever disappearing completely [24].

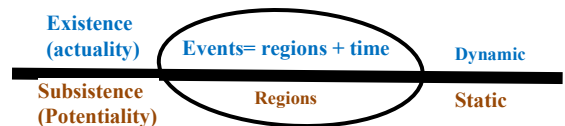

Fig. 2 Two levels TM modeling.

According to sources given in [25], "potentiality, as a mode of existence [TM reality] is a category far more relevant in an attempt to understand one's place in the world, than that of being. The latter is conditioned by the former." Currently, there is a paradigm shift from a linear logic of causality and narrative to a nonlinear, multifaceted world of potentiality that is being theorized as a crucial step to understanding contemporary world mechanisms [25].

### B. The TM model: Things are Machines and vice versa

Each of the two TM levels is associated with its *mode* (existing/subsisting) of thimacs. TM reality is built from a superpositioned, entangled, complementarity complex of thimacs. Thimacs are the wholes (entities, e.g., a stone; and processes, e.g., traffic) of worldview composition that can be built from subthimacs, thimacs, and so on.

Reality organizes itself according to these thimacs as (a) events (at the actuality level) and (2) regions (at the potentiality level). Note that members (thimacs) of the actual world are the same as the potential world, but they are different in their modes of reality: events or regions. We ignore magical or unreal (nonactual) entities at this research stage.

Entities of actuality are *events*, and corresponding entities of potentiality are *regions*. Events exist, and regions subsist. A region in substance becomes an event in existence when the region is injected with time. Concerning our current interest in propositional logic, the value of truth is a matter of how things are in actuality, and falsehood is in subsistence. Generally, what the proposition means and what the world is like are based on the TM representation: a thimac with actions: create, process, release, transfer, and receive (see Fig. 3). *Static* actions are regions that denote pure (timeless) 'forms' and *dynamic* actions (time-based) that are events.

According to these static (unchanging) and dynamic features, a thimac may be called a *thing* or *machine*. A thimac is a *machine* when it acts on other thimacs, and it is a *thing* when it is the object of actions by other thimacs (see Fig. 4). E.g., the thimac may *process* thimacs and may *be processed* by other thimacs.

A thimac may be constructed from a subset of the five generic actions. A *create* action is mandatory for any thimac since it denotes subsistence at the static level and existence at the dynamic level. A thing can be made, processed, released, transferred, and received. A machine is defined in terms of the generic action according to the structure shown in Fig. 3. For simplicity's sake, we assume that things are always **accepted** at their destination (see Fig. 3); therefore, we focus on the five red actions in the figure.

Additionally, the TM model includes *storage* and *triggering* (denoted by a **dashed arrow** in this paper's figures), which initiate a flow from one machine to another. Triggering transforms from one series of movements to another (e.g., electricity triggers heat generation).

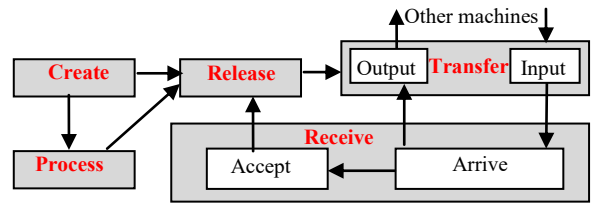

Fig. 3 Thinging machine

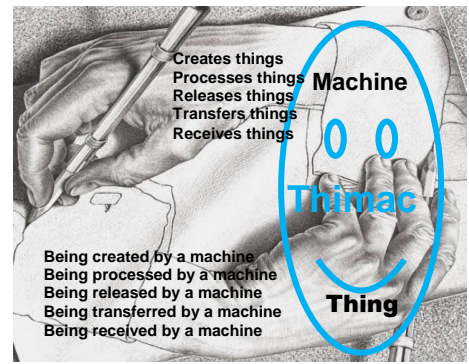

Fig. 4 A thimac can be a thing or machine– Escher's sketch

### C. Thimac as a Thing

A thimac may stand for a thing or a process (machine). At the actuality level, a thing is an event as much as a 'process' is an event. An event may be an existent thing or a process (something happening).

A thing is defined as what can be created, processed, released, transferred, or received. All things can flow (i.e., things that can be created, released, transferred, received, and processed). A thing flows from a machine through its *release* and *transfer* (output) to another, reaching this second machine's transfer (input) to settle in its receive. Flow refers to a transformation in the situations of a thing among creation, procession, releasing, transferring, and receiving.

### D. Thimac as a Machine

Machines are defined by Fig. 3 as their blueprint. This blueprint is a static 'form' at the potential level and becomes a dynamic event (region+time) at the dynamic level. The thimac *machine* executes five actions: *create*, *process*, *release*, *transfer,* and *receive*. Each static (outside time) action is a capacity to act and becomes a *generic event* when merged with time. Thimacs are *realized* by creating, processing, releasing, transferring, and receiving thimacs. Each thimac is affected by/affects the thimacs in contact with it through releasing, transferring and receiving.

A thimac's actions, as shown in Fig. 4, are described as follows.

*1)Arrive:* A thing arrives at a machine.
*2)Accept:* A thing enters the machine. For simplification, the arriving things are assumed to be *accepted* (see Fig. 3); therefore, *arrive* and *accept* combine actions into the *receive* action. Thus, the thing becomes inside the machine.

3) *Release:* A thing is ready for transfer outside the machine.

4) *Process:* A thing is changed, handled, and examined, but nothing new is generated.

5) *Transfer:* A thing crosses a boundary as input into or output from a machine.

6) *Create* A new thing revealed in a machine. The creation action refers to potential subsistence at the TM static level and actual existence at the TM dynamic level.

### E. Subsistence in Existence

Existence and subsistence are like a double-image impression (e.g., Rubin's vase), which is possible with a figure-ground perception. Creating an event does not change the region but rather mixes it with time to produce an event. Still, the region is in the interiority of the event without any change; hence, eventually, the event is reverted to a static region.

### III. TM MODELING EXAMPLE

Over the last few years, many TM modeling examples have been introduced (see the Google Scholar site). This paper provides an additional example selected to illustrate TM modeling while it has propositional logic components.

### A. Drawing a Conclusion from Premises

According to reference [18], logic can be seen in action around us. Consider the following example,

> In a restaurant, your Father ordered Fish, your Mother ordered Vegetarian, and you ordered Meat. Out of the kitchen comes a new person carrying the three plates. What will happen? The waiter asks the first question, says, "Who ordered the meat?" and puts that plate. Then he asks a second question, "Who has the fish?" and puts that plate. And then, without asking further, he knows he has to put the remaining plate in front of your Mother. [18]

The waiter draws a conclusion when the waiter puts the third plate without asking. The two answers received allow the waiter to infer where the third dish must go. This inference schema can be represented as follows (using F for "fish," M for "meat," and V for "vegetarian"): F or V or M, not M, not F ⇒ V.

### B. TM static Model

Fig. 5 shows the static TM diagram of this example. To explain this diagram, we isolate its structure in Fig. 6.

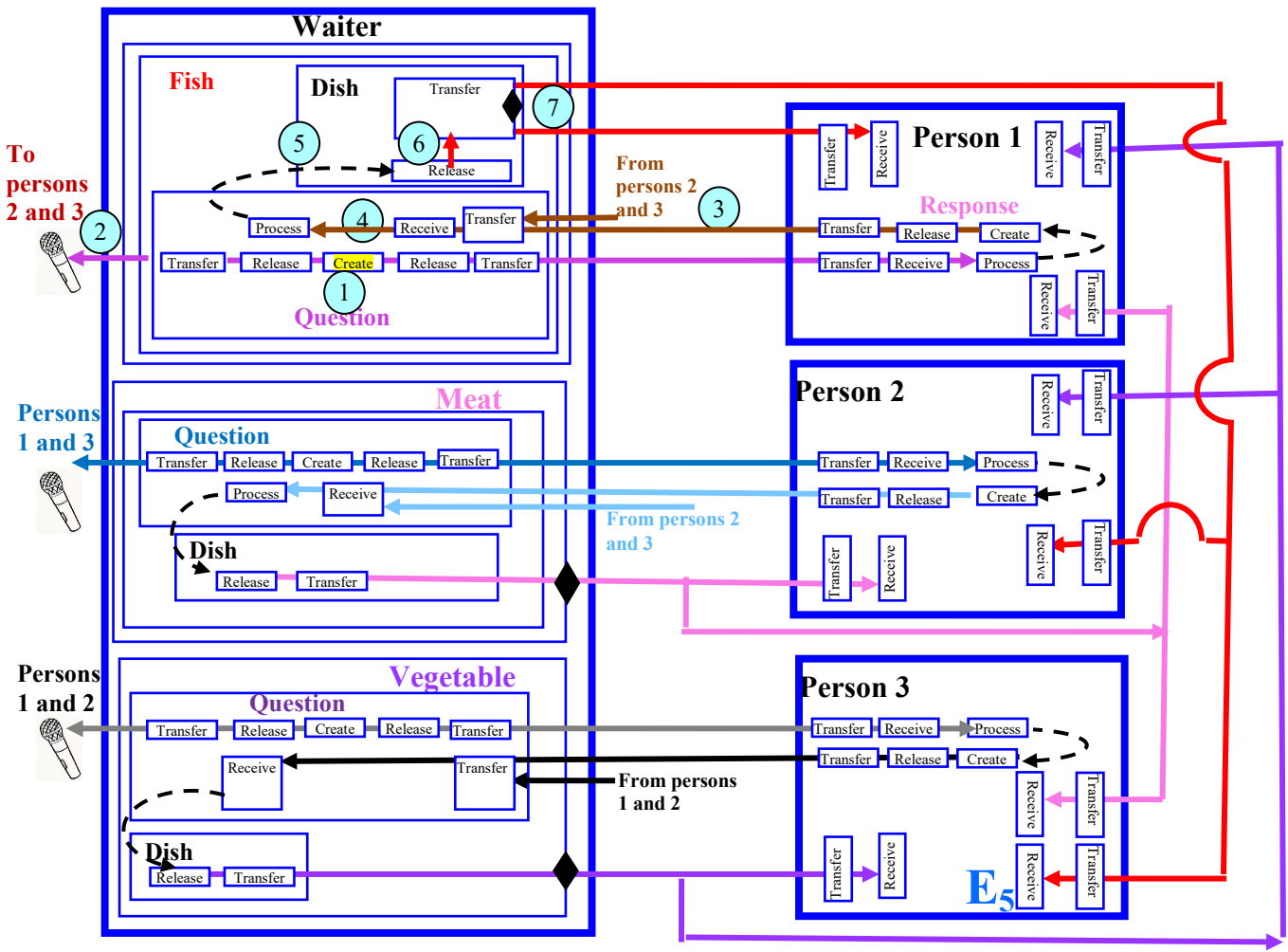

Fig. 5 The static TM model

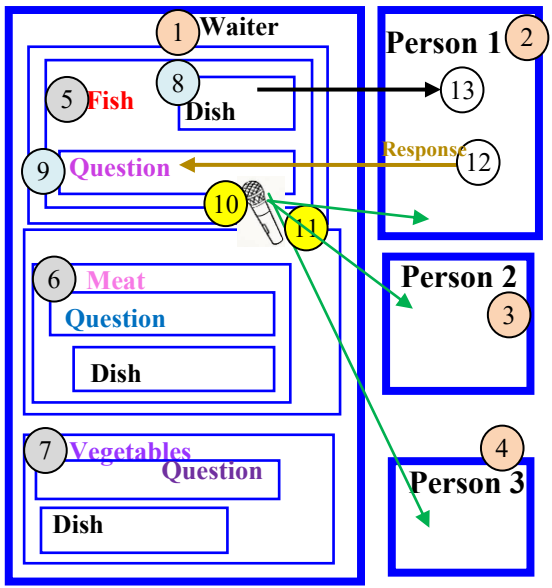

Fig. 6 The general structure of the static TM model.

Fig. 6 includes the waiter on the left (the circled number 1) and the three persons waiting for their dishes on the right. Accordingly,

- The waiter has three dishes: fish (5), meat (6) and vegetable (7).
- Each dish has two related thimacs (indicated by boxes). For example, in the fish box are the fish dish (8) and the question about who ordered fish (9).

Similarly, each box contains two boxes for the meat and vegetables: the dish and the question about who ordered it.

- The figure also shows that the fish question is directed toward all three persons (for the fish order, this is symbolized by the microphone (10)). The arrows (11) indicate the flow of the fish question to the three persons.
- Suppose the first person responds (12) that he/she is the one who ordered the fish. Then, that dish is delivered to the first person (13).

Note the simplicity of the whole diagram, which involves repeating subdiagrams of the type of dish (e.g., fish) no matter how many kinds of dishes are present. Also, the same subdiagram of communication with a person is repeated, no matter how many persons are present.

Now, we return to describe the static model of Fig. 5.

- Assuming that the waiter has started to serve the first dish, he asks which person has ordered fish (1), and the question flows to the three persons (2).
- He receives the responses (3, 4).
- The response triggers (5) the release (6) of the fish dish to the responded person (7). The black diamond (7) is a diagrammatic simplification that indicates that the fish dish may go to any of the three persons depending on the processing of the response.
- The same description can be applied to the other two cases: meat and vegetable dishes.

Fig. 5 can be simplified by assuming the arrows indicate the flow direction. Hence, release, transfer, and receive can be removed from the diagram, resulting in Fig. 7.

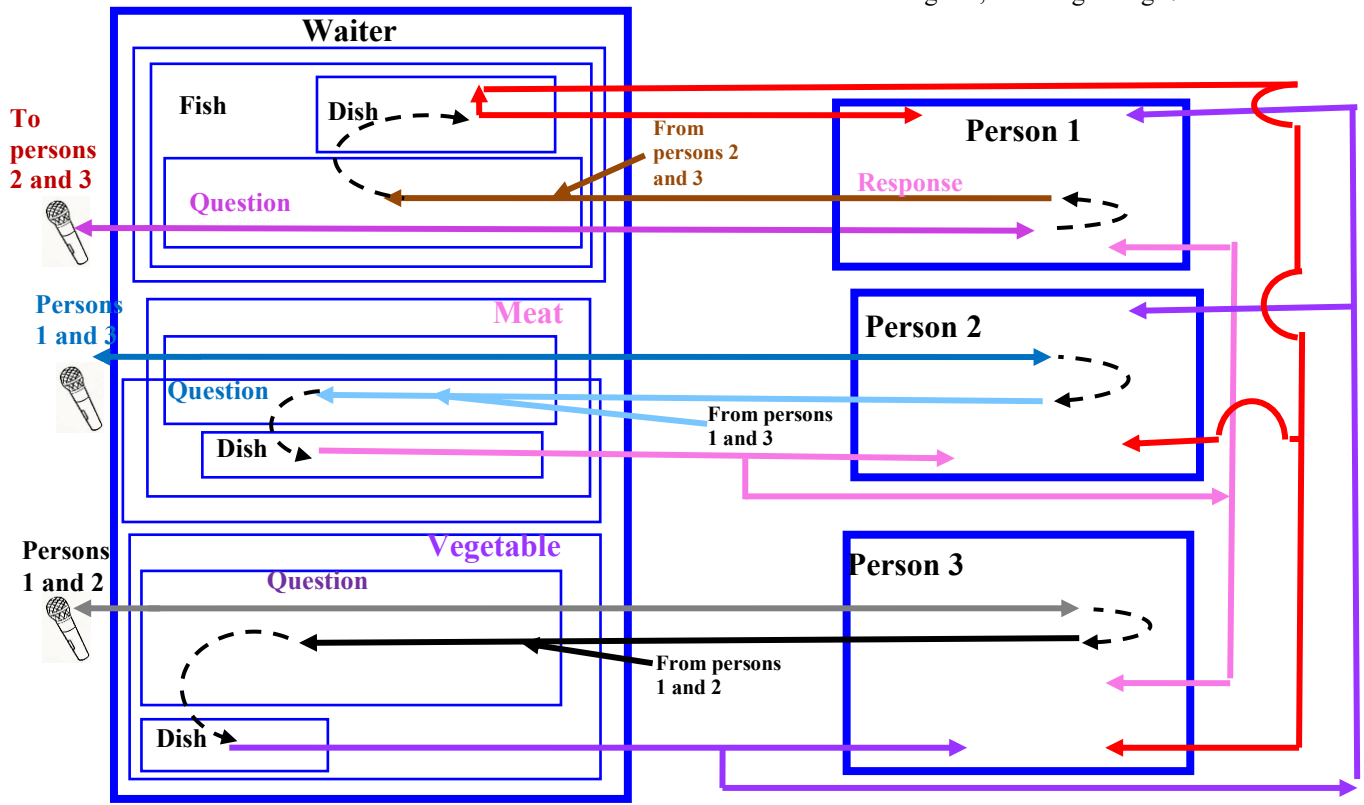

Fig. 7 Simplified static TM model.

*C. Dynamic model*

The dynamic model is built upon the notion of event. An event is a region (static thimac) plus time. The region is a subdiagram of the static model. For example, Fig. 8 shows the event where *The waiter serves the meat dish to the second person*.

For simplification's sake, the event will be represented by its region, as shown in Fig. 9. Accordingly, Fig. 10 shows the dynamic model corresponding to the static model of this waiter decisions example. It includes the following events.

$E_1$: The waiter asks who ordered the fish.

$E_2$: A response is received.

$E_3$: The fish dish is delivered to the responsive first person.

$E_4$: The fish dish is delivered to the responsive second person.

$E_5$: The fish dish is delivered to the responsive third person.

$E_6$: The waiter asks who ordered the **meat**.

$E_7$: A response is received.

$E_8$: The meat dish is delivered to the first person.

$E_9$: The meat dish is delivered to the second person.

$E_{10}$: The meat dish is delivered to the third person.

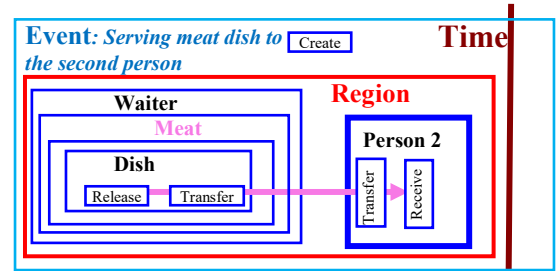

Fig. 8 The event *The waiter serves the meat dish to the second person.*

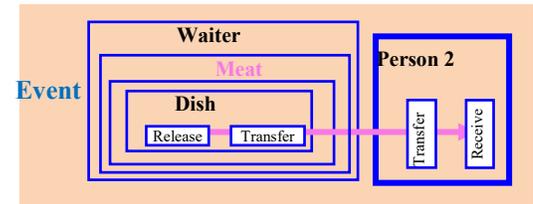

Fig. 9 Simplification of the event *The waiter serves the meat dish to the second person.*

$E_{11}$: The waiter asks who ordered the vegetable.

$E_{12}$: A response is received.

$E_{13}$: The vegetable dish is delivered to the first person.

$E_{14}$: The vegetable dish is delivered to the second person.

$E_{15}$: The vegetable dish is delivered to the third person.

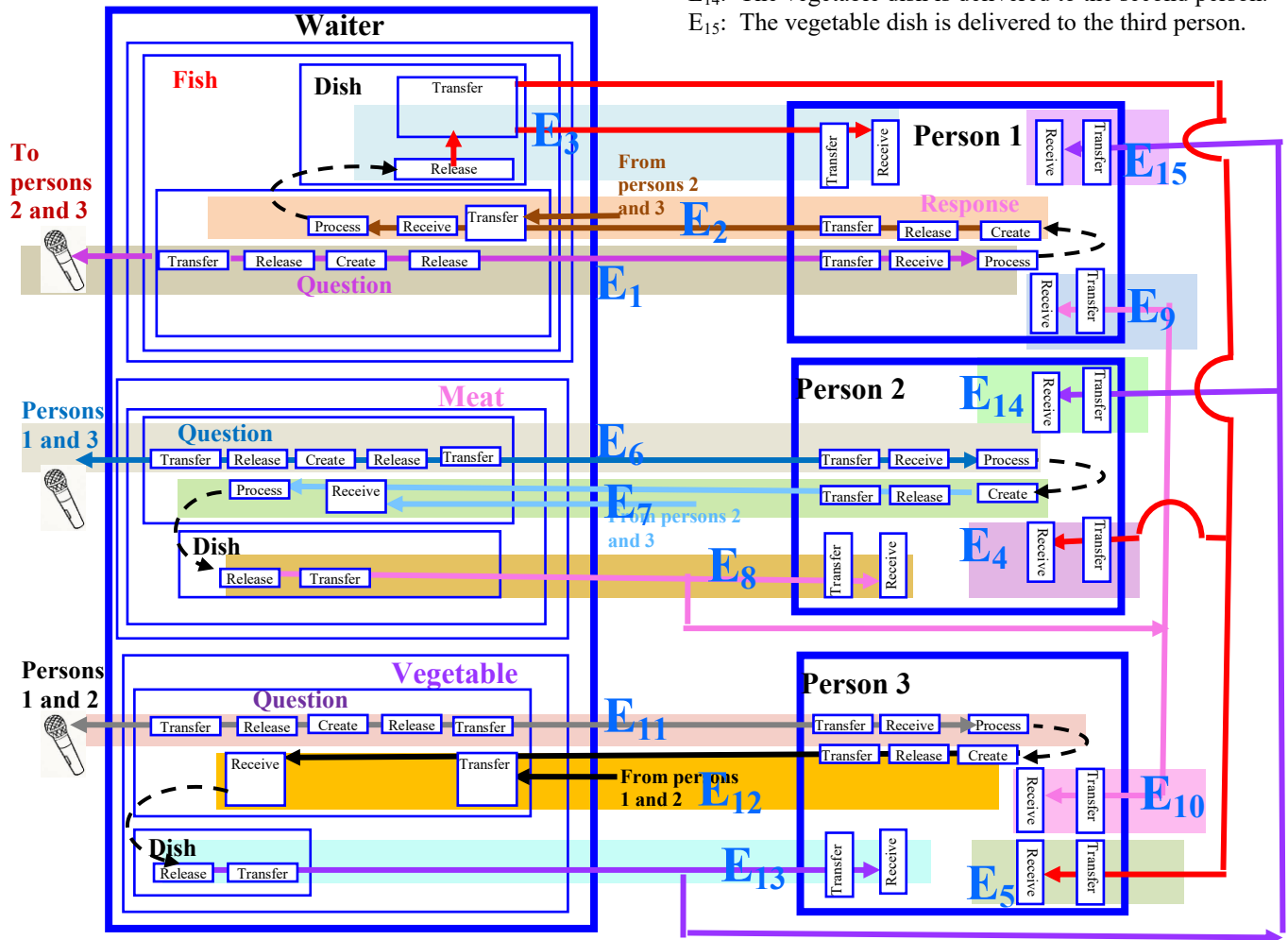

Fig. 10 Dynamic model

Let $E_0$ be the event of delivering one dish by the waiter. Fig. 11 shows the chronology of events of the waiter model. All 'legal' sequences of events can be identified from Fig. 11. For example, Fig. 12 shows the sequence:

$E_0 \rightarrow$ (fish) $E_1 \rightarrow E_2 \rightarrow E_5 \rightarrow E_0 \rightarrow$ (meat) $E_6 \rightarrow E_7 \rightarrow E_8 \rightarrow E_0 \rightarrow$ (vegetable) $E_{13}$ as follows.

$E_0$: Serving the fish dish.
$E_1$: The waiter asks who ordered the fish.
$E_2$: A response is received.

$E_3$: The waiter gives the fish dish to the first person (Apparently, the first person responded that he ordered fish).
$E_0$: Serving the meat dish.
$E_6$: The waiter asks who ordered the meat.
$E_7$: A response is received.
$E_8$: The waiter gives the meat dish to the second person (Apparently, the second person responded that he ordered meat).
$E_0$: Serving the vegetable dish.
$E_{13}$: The waiter gives the fish dish to the third person.

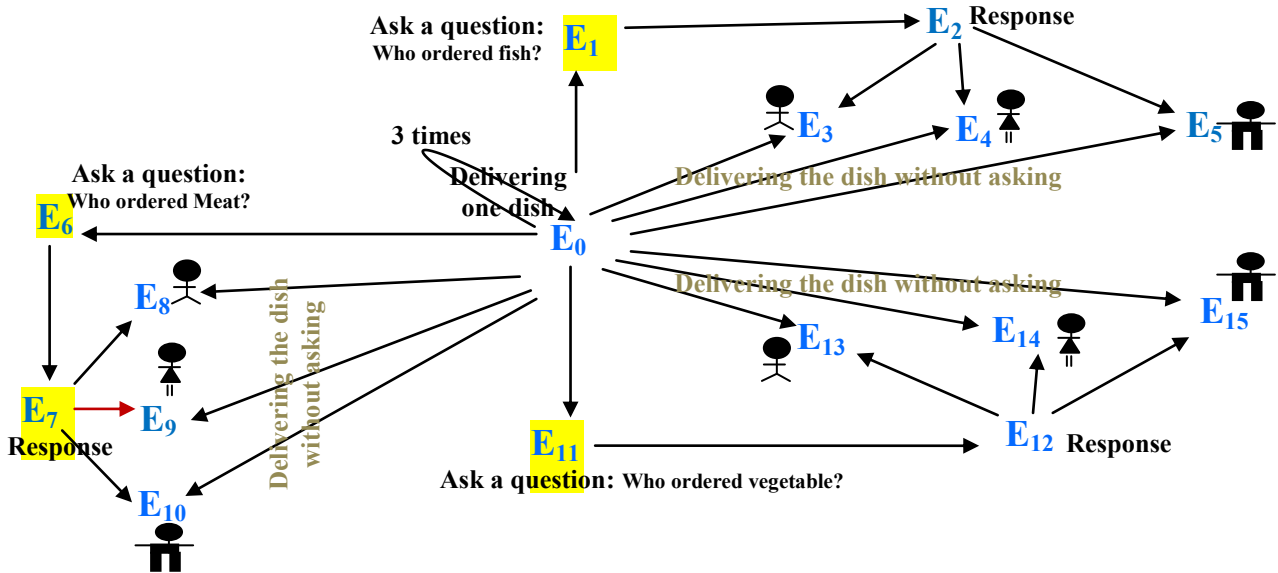

Fig. 11 The chronology of events of the waiter model

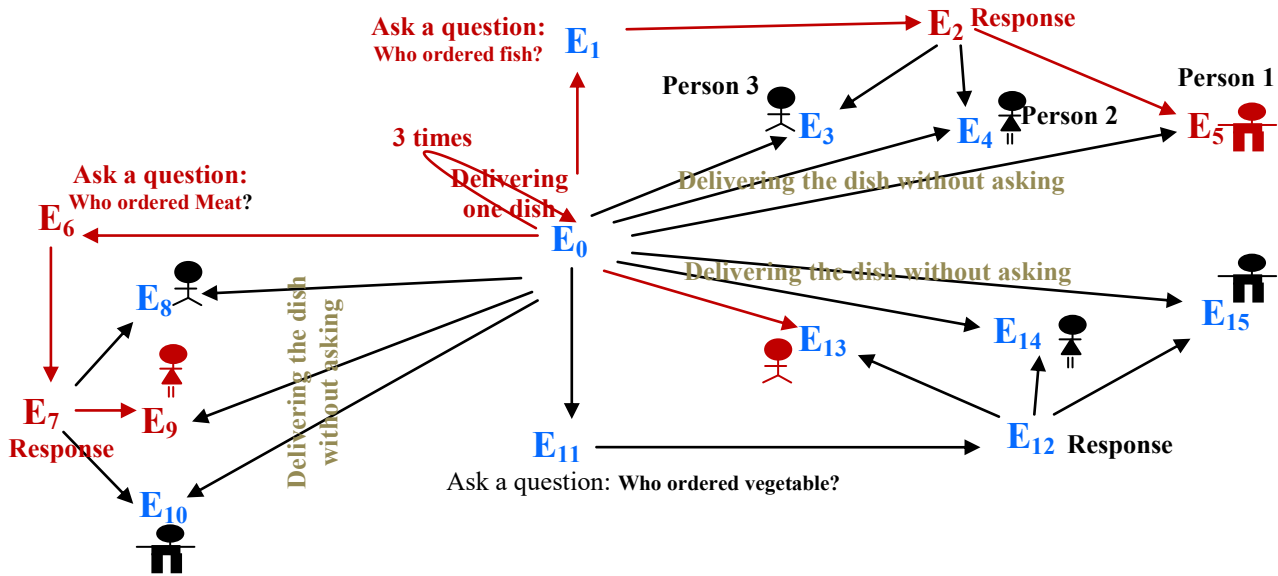

Fig. 12 The chronology of events of the waiter model starting with the question who ordered fish then who ordered meat (red arrows)

## IV. PROPOSITIONAL LOGIC

The exploration of the connections between the reality and logical structures employed involves describing or discovering reality. According to Prior (quoted in [26]), logic "is not primarily about language, but about the real world. [...] Formalism, i.e., the theory that Logic is just about symbols and not things, is false. Nevertheless, it is important to 'formalize' as much as possible, i.e., to state truths about things in a rigorous language with a known and explicit structure." In this section, we explore the logic-reality relationship using the TM model.

### A. Propositions Types: regions and events

Propositions are regarded as performing vital roles in theories of natural language, logic, and cognition [27]. They are often introduced in terms such as the meanings of sentences, bearers of truth and falsity, and objects of belief and entities related by the relation of logical consequence [28]. In TM, a proposition is a thimac. This is a sweeping proclamation since it indicates that a proposition is not only a *thing* (e.g., *Aristotle is human*) but also a *machine/process* (e.g., *Aristotle eats food*). Furthermore, this thing or machine has a static 'form' (called the region) and a dynamic actuality (called an event).

Consider the following introductory samples of propositions and their TM representations.

*Example*: *Felix is a cat* is a region (potentiality for actualization). It is a thimac that can be processed, released, transferred, and received. It specifies a member of the set cat and has the subthimac named create *Felix*. Additionally, *Felix is a cat*, a machine that includes actions such as receiving and processing food. In its regional form, this machine may be actualized as an existing cat with possible actions.

*Felix is a cat* represented in TM as a region and event in Fig. 13. No such distinction is recognized when *Felix is a cat* is declared as a proposition.

*Example*: *Sam is sad* is true [29]. It can be represented in TM as a region with property *sad*. Suppose that the property of being sad has the form ∃x Property (x) & x = being sad, i.e., the *property* is an independent thimac. According to [29], some account must be given of the fact that grasp of the expressions, *The property of being sad* and *The proposition that Sam is sad.* Fig. 14 shows the TM representation of *The property of being sad* where, for simplicity sake x, assumes to exists. At the dynamic level, we see two events: (1) the property sadness is an actual property, and (2) Sadness flows to x (Sam).

Accordingly, a logical proposition in TM can be of two types: regions or events. Propositional logic does not distinguish between these types of propositions. Typically, in Logic, events are contrasted with *facts* characterized by features of abstractness and temporality, e.g., the event of Caesar's death took place in Rome in 44 BCE, but that Caesar died is a fact here as it is in Rome today, as in 44 BCE.

In TM, *Caesar's death is an event because it is stated in the past tense, along with Caesar's death, which took place in Rome in 44 BCE,* which gives a specific time. However, *Caesar is a dead person* is a region and a living person in a region. The *fact* notion in logic does not distinguish events and areas. It may be speculated that for every event, there is a companion fact, viz., the fact that the event took place [30]; however, in TM, for every event, there is a companion of static region that the event has realized.

The centrality of events is not a new idea. For example, *event semantics* claims that clauses in natural languages describe events [31]. Event descriptions are formulas like ∃e[P(e)], where e is a variable over events, and P stands for a predicate that is either simple or complex. However, events are objective individuals with a distinctive relation to time. "Unlike facts, events take place at or in times. *Obama won his first presidential election.* The event took place in 2008, but the fact did not, and is the same now as it was ten years ago." [31]. Instead of such notions as events/facts and events/descriptions, TM provides the duality of event vs. region.

Note that a region is a more general notion than a proposition. If R is the two-place relation that p asserts to hold between *a* and *b*, then p is true if and only if *a and b* stand to each other in the relation R, while p is false if and only if it is not the case that a and b stand in the relation R. In TM, R is a thimac involving *a* and *b*. R is valid if it is a (higher level) event, while it is false if it is a 'complex region' that can be mapped to actuality. Thus, truth is event-sizing a region. P is valid, which means that P is an event. For example, *Jones' report was more true than Roberts'* [32] expresses that Jones's report includes more events than Roberts.,

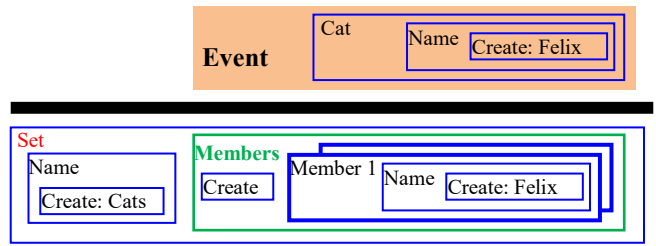

Fig. 13 Region and event of *Felix is a cat*

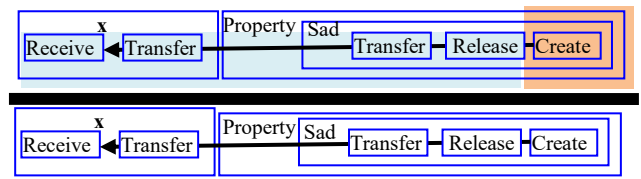

Fig. 14 Region and event of *∃x Property (x) & x = being sad.*

*B. Negative Propositions: Circumventing 'not Existence' is 'not Potentiality'*

As stated previously, true propositions are sometimes called facts. This brings up issues related to *negative facts* (given that there are true negative propositions) in addition to atomic facts and whether propositions had ordinary objects or only concepts as constituents [28].

The negative proposition is a significant issue when representing propositional logic in TM modeling. It is claimed that negation appears to be a primitive element of our processes of thinking and knowing anything [22]. In the Aristotelian tradition, negations can express a contradiction, a contrariety, or a subcontrariety [22]. Additionally, according to Wittgenstein (as quoted by [22]), "The positive sentence must presuppose the existence of the negative sentence."

To analyze the negative proposition in this situation, we first give the following general explicit picture of the totality of the two levels of the TM model. In such a picture, 'Not existence' corresponds to potentiality, 'Not potentiality' corresponds to what we will call 'absent event,' and other cases of (non) potentiality that are outside the two-level TM model are viewed as a 'hole' in the level of existence.

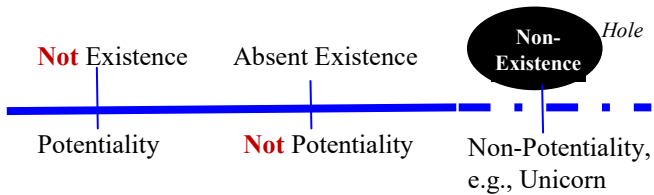

Fig. 15 illustrates the non-existence and potentiality using p to denote a sample proposition. Fig. 15 shows permitted mapping between regions and events. At the dynamic level, p̄ (circle 1, events will be darkened) denotes p. ¬p (2) denotes the subsisting region p. In this case, **¬p** means a non-existing event; however, its non-existence does not mean losing its potentiality. ¬p (not event) 'means' a subsisting p (3, region).

On the other hand, a subsisting region p (3, not darkened at the static level) may be negated as a negated region ¬p (4) is actualized as an absent event. Note that p is assumed to have potentiality. It can be actualized as the *absent event* p (non-darkened at the dynamic level). The absence of an absent event is purely temporal, i.e., it can be an actual event earlier or later. In philosophical language, an absence in this context means potentiality conserves its nature (staticity) in actuality. An absent event can be a sub-event of an actual event, e.g., a broken actual car has an absent functionality.

Absent event p semantics is explained in Fig. 16. The negativity of region ¬p means *de-creating* region p at the actuality level. Not that, in Fig. 16, an event is created with its region being de-created. De-create refers to 'performing' the region without realizing it. The de-created region occupies the same time as the actual event but without 'activating' it. ' It is a form of imitating an event or performing it in a mirror. The event does not exist, but everything else (e.g., time, sub-events times, intensity, etc.) is in the mirror. The absent event function preserves event-ness (time nature) with a changeless region.

Fig. 17 illustrates these ideas further. E and R denote an event and region, respectively. The event is represented as a 'function' of time and region, which are simplified as E, ¬E, and E.

A missing event (not in the dynamic level) will be left out of the dynamic model as a 'hole' in existence. If absent or missing events border each other, they will be separated by dotted boxes or lines in the TM diagram.

This idea of entwining actuality/potentiality in absent existence may be related to Aristotle's dualistic ontological system (which distinguishes between actual and potential being). Virtuality is the third form of being that can complement this system [20].

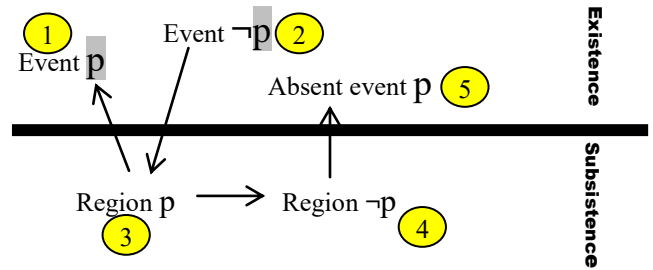

Fig. 15 Mapping between the TM regions and events including all permitted negations.

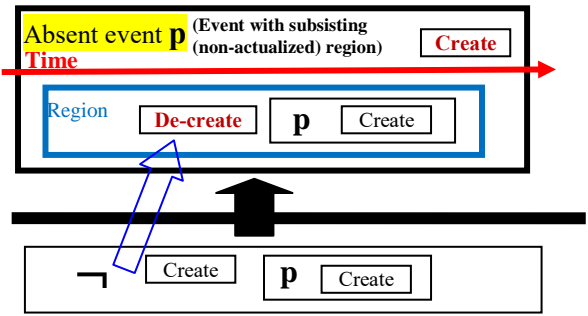

Fig. 16 Negative region and its absent event

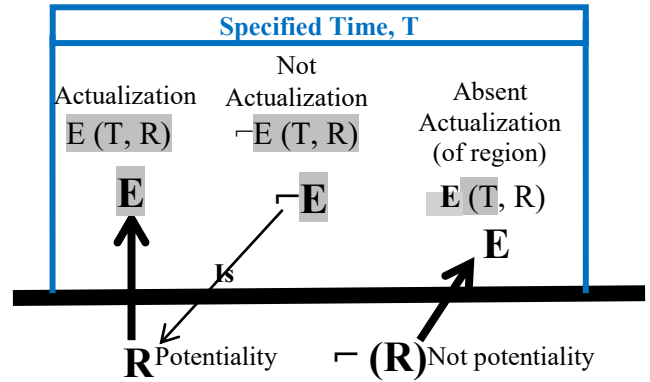

Fig. 17 Mapping between the TM regions and events. Darkening implied actualization.

According to [20], "In the virtual form of being, actuality and potentiality are inseparably intertwined. Virtuality is potentiality considered together with its actualization. In this view, virtuality is a reality with a measure, which has no absolute character but a relative nature."

## C.   Types of Negativity: examples

Propositional logic does not distinguish between types of negativity at the static level vs. the dynamic level. Negativity at the static level, assuming potentiality, is the absence of the corresponding event. According to [33], "If I decide not to pick up my friend at the airport, I perform an act of *omitting* to pick him up" (Italic added). *Omission* of a thimac is a *mode* of the event (of actual thimac) left undone and an activation failure.

Consider negations that express genuine oppositions. *Ammonia is not an acid* [22]. From the TM point of view, *Ammonia is an acid* and is not a region; hence, it cannot be actualized. *Ammonia is not an acid*, which is not the negation of Ammonia. It is an acid, just as not a square circle, which is not the negation of a *square circle*. Both of them do not exist in reality, i.e., not actualized. *Ammonia is not an acid*; it is a denial of existence (hole).

Consider [22]'s example of "Mr. A is in the room." In TM, *Mr. A is in the room,* is a potentiality, and to make it an event, we may add time as *Mr. A is in the room now. Mr. A is not in the room* is a negation of a region; hence, it refers to the absence of Mr. A in time now. However, *Mr. A is not in the room **now**, which negates the event,* is the region *Mr. A is in the room.*

# V.   LOGICAL CONSTANTS

Logical constants (e.g., ∧ and ∨) relate two thimacs to make a new thimac from them. Such logical structures are shared across application domains.

## A.   Disjunction

Events can be complex or super events. Just as to say that propositions are structured is to say that they are complex propositions. Note that each thimac in (A∨B) could be realizable (be an event) without the ∨ thimac being realizable. Hence, a super thimac can be more than a relation because some parts may act independently from the relation.

TM differentiates between regions and events in logical A∨B can be viewed as,

- Logical A and logical B are regions (potential thimacs) at the static level.
- Logical A or logical B could be events; in this case, they will be represented in TM as A and B at the dynamic level.

Thus, A∨B is represented in the event level as A ⋮ B ⋮ AB using a vertical dotted line to denote alternatives. At the static level, A∨B is defined as the supper thimac ∨ that contains the two regions, Such a TM consideration has immediate consequences related to classical contradictions.

*Example*: Given A∨B and ¬A, then logic implies B. According to [34], from a contradiction, anything follows—E.g., From A and not A, to show B.
1. A (A in TM because of the true value)
2. A OR B (A OR B in TM because of the true value)
3. 'NOT A' (In TM, a negation of a region)
4. B

Note that 'NOT A' refers to an *absent event* since **it is true** 'NOT A' is such an event. Hence, instead of the traditional logic notion of "it is not the case that," a negative (region) proposition means that the corresponding event is absent (from the previous section's interpretation). While A OR B (2 above) implies actualized existence (by their truthness), it (i.e., A OR B) is not true when 'NOT A' is interpreted as an absent event. That is, 'NOT A' does not imply B. The disjunction (2 above) introduction cannot include the absent event of 'NOT A,' i.e., absent A does not imply *absent A* OR B. Thus, (4) above cannot be deduced from (2) and (3). The TM distinction between actual and absent events blocks the deduction that leads to (4)—the same conclusion when we assume NOT A.

## B.   Not Potentiality and Absent Existence

Absent events are not mere absences. They are events with de-created regions, i.e., occurrences with a changeless regions. Such an absent event duality of dynamic time and static region may be related to Lupasco's conception of energy, which "possess a logic that is not a classic logic nor any other based on a principle of pure non-contradiction, since energy implies a contradictory duality in its nature, structure, and function [TM region-ness and event-ness]" [16]. Structure and function, here, can reflect TM region-ness and event-ness, respectively. According to Lupasco, as quoted in [16], "The contradictory logic of energy is a real logic, that is, a science of logical facts and operations." As formulated by Lupasco, the critical postulate is that every event *e* is always associated with an anti-event *none such that the actualization of e entails the potentization of none* and vice versa, "without ever disappearing completely. This is formalized into three fundamental axioms of classical logic that this paper will not discuss since this topic needs its research treatment.

The absence of events should be viewed as events in their own right [33]. In TM, the absence of an event at the dynamic level requires a region in the model that preserves the event's absence. *Randy omits to pick up milk [33], is* an absent event of the region where Randy picks *up milk.* In such an absent action, Randy (deliberately) exercised an event just as much as Randy would have if he had done the event of picking *up milk (this morning).* Both types of events have their time slot. The negative region combines with time to reserve its 'position' as a type of event.

The proposition *Socrates is not wise* affirms that *actual wise* (event) is absent in *Socrates (event)*. If *Socrates* does not exist, then *Socrates* is a region with the potentiality of being an event, and *not-wise* is the non-potentiality of actuality (absent event).

According to [35], "negative properties did not seem particularly natural and perhaps should be excluded from our ontology on that basis." In TM, a property, such as being a non-red thing, is an absence of redness in the thing. At the static level, the region (diagram) has the property of redness. The negation of this region is due to the absence of the corresponding event. Such 'features' as non-red deserve a place in reality at the subsistence and existence level *except* when the feature is impossible to actualize, e.g., non-square circle (See discussion in [35]). A red thing is a potentiality and actuality, and region negation is absent events. A square circle thing has no potentiality and, thus, has no actuality; hence, it is excluded from both TM modes of reality, called previously 'hole.'

Consider the distinction between *being not green and not being green*. The negative logical expression *x is not green* can be understood in TM, about actuality, in the following ways,

- *x is not* green, which is equivalent to *x is* green. This means that x has the potential to be green, but it is currently not green (e.g., chameleon).
- *x is not* green, which means the same as *x is green* (potentiality).
- *x is not green, which* means x is actualized, but its greenness is not actualized (absent event).
- *x is not green, which* means the whole thimac is not actualized.

On the other hand, the negative logical expression, *not x is green* or, more precisely, *not (x is green),* can be understood in TM, concerning actuality, in the following ways,

- *Not (x is green)*, i.e., not event, refers to the potentiality *x is green.*
- *Not (x is green) refers to the actuality spot x is green as an absent event.*

TM events unite positive regions (potentialities) and negative regions as events, actual and absent, respectively. The logical formula (p∨¬p) states that every proposition p is either true or false. In TM, it depends on whether p is a region or event. If p is an event, then ¬p refers to its region p. If p is a region, then ¬p denotes an absent event, assuming that p is realizable.

### C. Conjunction

Similar analysis can be applied to Aristotle's most certain of all principles that contradictory statements are not at the same time true: A∧¬A. In TM, if A is an event, then A in the level of the events and its region A (e.g., corresponds to logical ¬A). If A is a region, then ¬A is an absent event. So logical A∧¬A can denote two TM situations.

- A∧¬A at the events level (actual and absent event simultaneously) is not permitted.

- A∧¬A (where A and ¬A are at the static level) is permitted since ¬A may be realized as an absent event.

A logical issue here is that standard objections to the view that omissions are causes, is that if I caused the death of my neighbor*'s* plant by failing to water it, then so did everyone else (since no one watered the plant, everyone must have committed an act of omission) [36]. According to [36], it can be claimed that "so did my plumber's not watering my neighbor's plant." In TM, *my plumber's not watering* is a negation of a region with the absent event, p (i.e., I was the one who promised, and my neighbor did not ask my plumber to water her plant); similarly, in the claim that *If I had watered my neighbor's plant* (absent event), *it would not have died* (negative event: a region). According to [36], omissions (facts or proposition-like entities) are not events, and causation relates to events. In TM, the event of different modes (actual, negative, and absent) unifies causal explanations.

### VI. CASE STUDY

According to [19], the requirements for borrowing and returning a specific book from a school library can be formalized in propositional logic. The book can be in any of the following four states: *on_stack, on_reserve, on_loan,* and *returned.* The propositions model these as,

- S—the book is on the stacks
- R—the book is on reserve
- L—the book is on loan
- Q—the book is returned

Constraints are specified as,

1. The book can be in only one of the three states: S, R, and L.
2. If the book is returned, it is on the stacks or reserve.

The propositional logic formulas for the constraints are:

1. S ⇒ ¬ (R∨L)
2. R ⇒ ¬ (S∨L)
3. L ⇒ ¬ (S∨R)
4. Q ⇒ S∨R

Suppose we want to prove that the statement "if a book is on loan, then it is not returned" is a consequence of the requirements.

### A. TM Static Model

Fig. 18 shows TM conceptualizing the problem as a static model. In Fig. 18, the diagram includes the stacks and reserve (circles 1 and 2 in the figure) where the books are located. Book Loan involves the following processes.

- Taking the book from the stacks (3).
- Processing it to register it in the loaned books record (4).
- The book is being taken outside the library (5).
  Returning the book involves the following processes.
- Bringing the book to the library (6).
- Removing the book from the record of loaned books (7).
- Sending the book to stacks (8) or reserves (9).

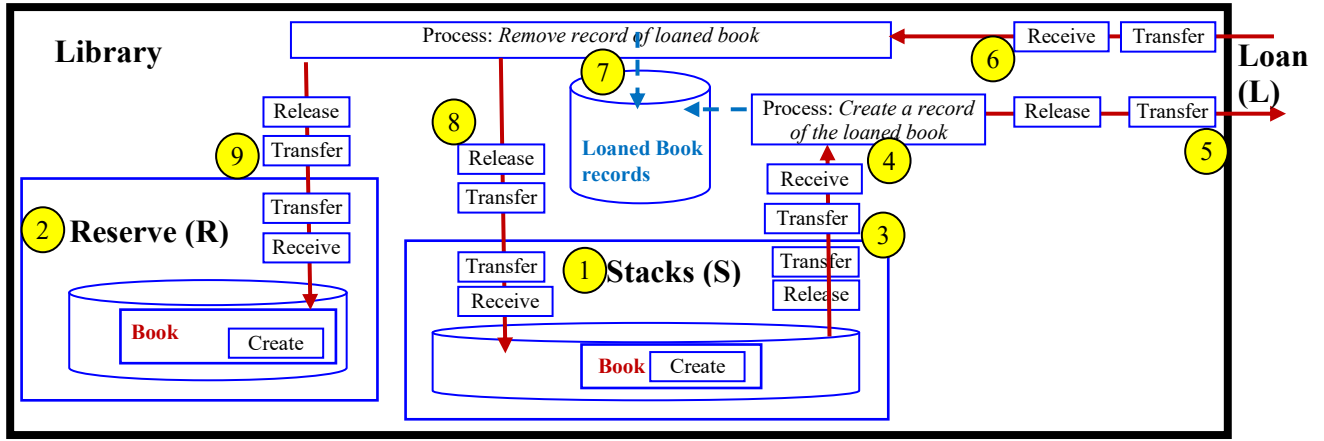

Fig. 18 The static TM model of the library

## B. Dynamic Model

Fig. 19 shows the events model as follows. Note that a darkened E denotes an event, and a non-darkened event denotes an absent event. Also note that a negative event is a region, i.e., $\neg E_i$ (dynamic level) is $E_i$ (static level).

$E_1$: The book in the stocks.
$E_2$: The book has been moved out of stock.
$E_3$: The book is processed to trigger, recorded as a loaned book, and taken outside the library.
$E_4$: The loaned book is returned to the library and processed to remove it from the list of loaned books.
$E_5$: The book is put in the stacks.
$E_6$: The book is put in the reserve.

Fig. 20 shows a typical chronology of events of borrowing a book.

## C. Propositions

The propositions S, R, L, and Q are regions (subdiagrams) on the static model, as shown in Fig. 21. Fig. 22 shows S, R, L, and Q as events. We show here how to model ¬S, ¬R, ¬L, and ¬Q in order to model the constraints.

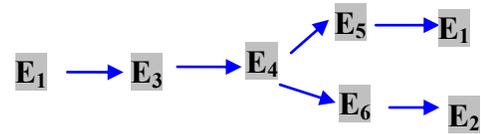

Fig. 20 The chronology of events in the library example

The given propositions are S (the book is on the stacks), R (the book is on reserve), L (the book is on loan), and Q (the book is returned).

1.  $S \Rightarrow \neg(R \lor L)$: This constraint is interpreted as the event $\boxed{S}$ implies the absence of events $\boxed{R}$ and $\boxed{L}$. The absence of $\boxed{R}$ and $\boxed{L}$ is indicated at the dynamic level as (non-darkened) R and L. The chronology of events of this constraint is presenting $\boxed{S}$, R, and L vertically to indicate one event includes the simultaneous occurrences as,

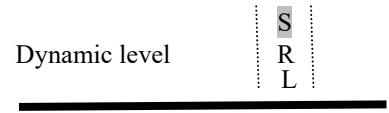

This expresses diagrammatically that S is realized only with the absent events R and L, as shown in Fig. 23.
.

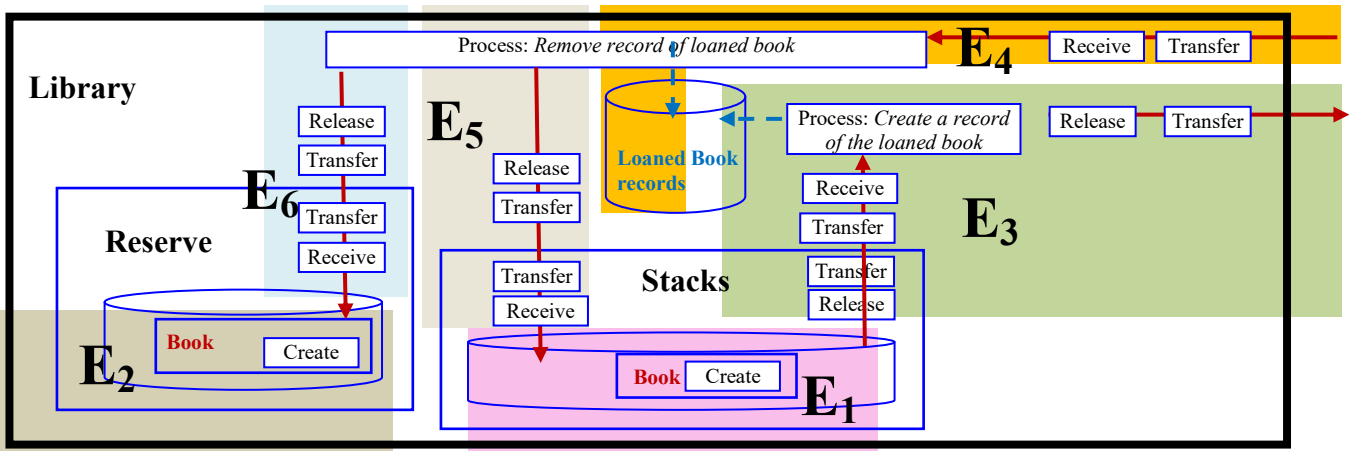

Fig. 19 The events in a typical book borrowing process.

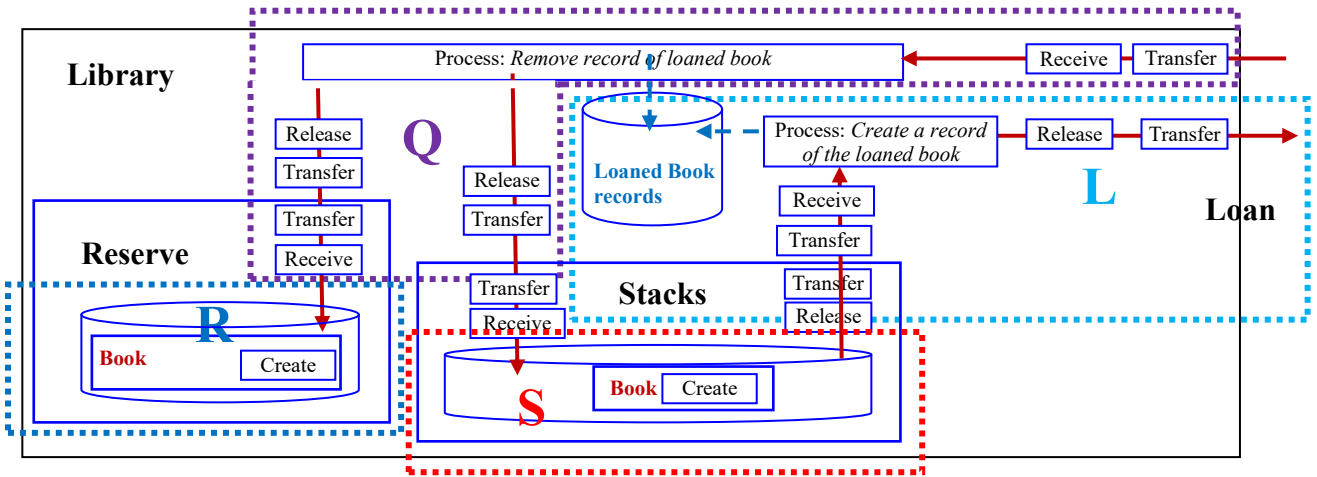

Fig. 21 The constraints S, R, L and Q

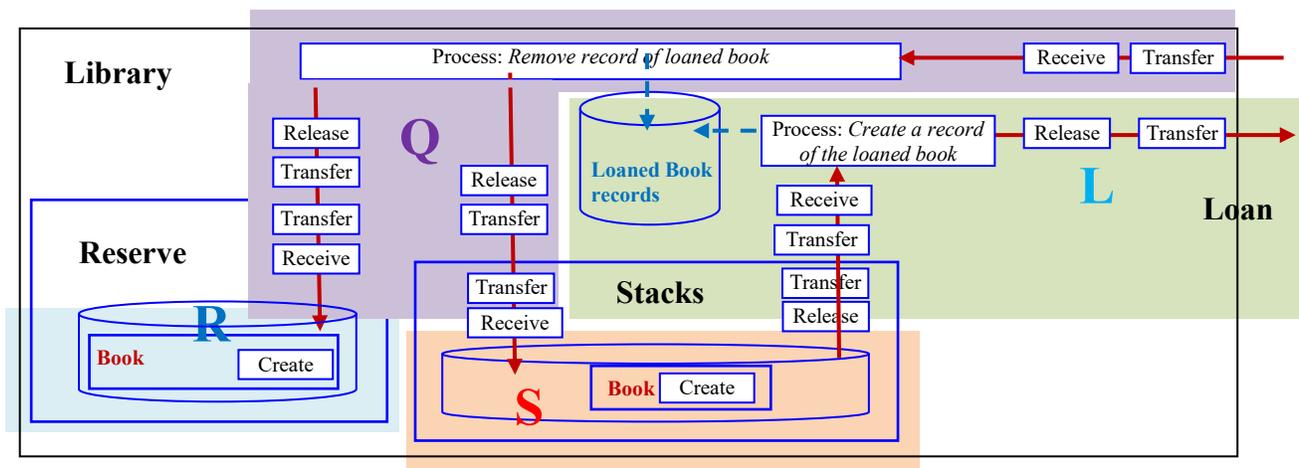

Fig. 22 The events S, R, L and Q

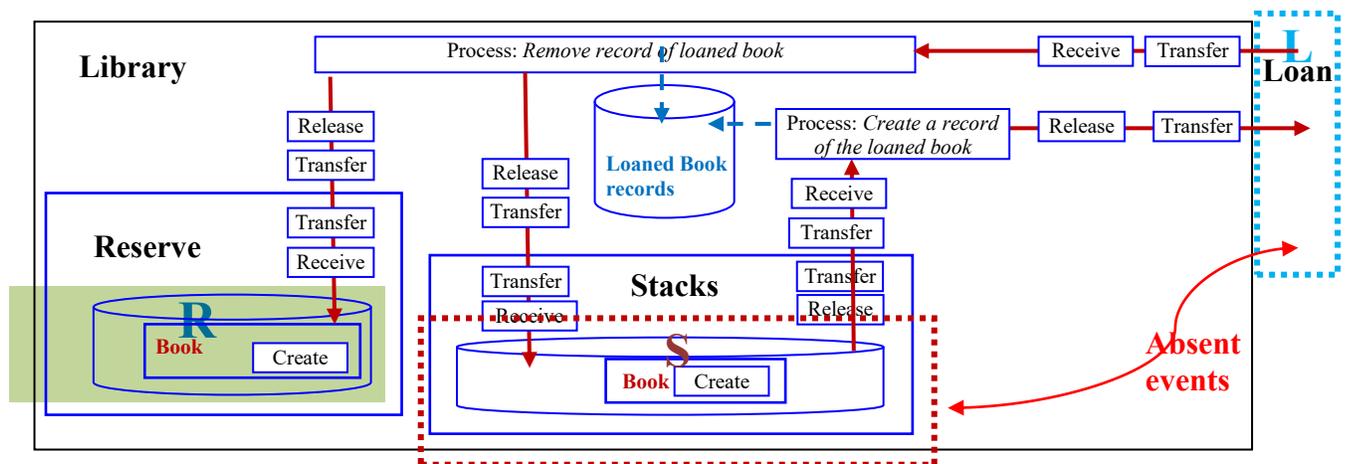

Fig. 23 Dynamic model of the constraints S ⇒ ¬(R ∨ L)

2. R ⇒ ¬(S ∨ L) and L ⇒ ¬(S ∨ R) are modelled in a similar way to S ⇒ ¬(R ∨ L).

3. Q ⇒ S ∨ R: This constraint is TM interpreted as Q, S, and R as shown in Fig. 24. The chronology of events is (Q ─ → S ∨ R), where the dashed arrow indicates triggering, not sequence, i.e., not after finishing Q. The ∨ indicated branching into two ∨-labelled arrows one leads to S, and the other to R interpreted as follow,

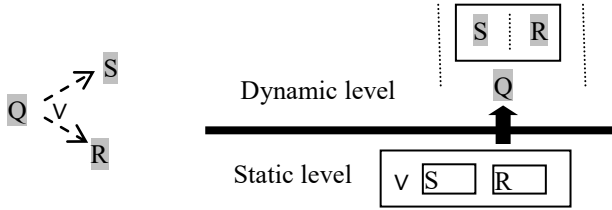

According to [19], suppose we want to prove that the statement "if a book is on loan, then it is not returned" is a consequence of the requirements.

In [19], this is achieved by including the negation of the formula L→¬ Q. Resolving ¬ (¬L ∨ ¬ Q) with the premise to get the two clauses (c4) L and (c5) Q to derive the empty clause NIL

Fig. 25 shows the TM representation of L and Q (absent event). From the chronology of events (Fig. 20), we can produce the new chronology of events, as shown in Fig. 26, as a version of the original chronology of events in Fig. 20. If L and Q are absent events (the book has not returned), then the sequence of events obviously does not lead to Q.

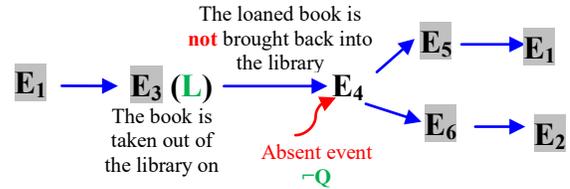

Fig. 26 The TM model of *If a book is on loan then it is not returned*

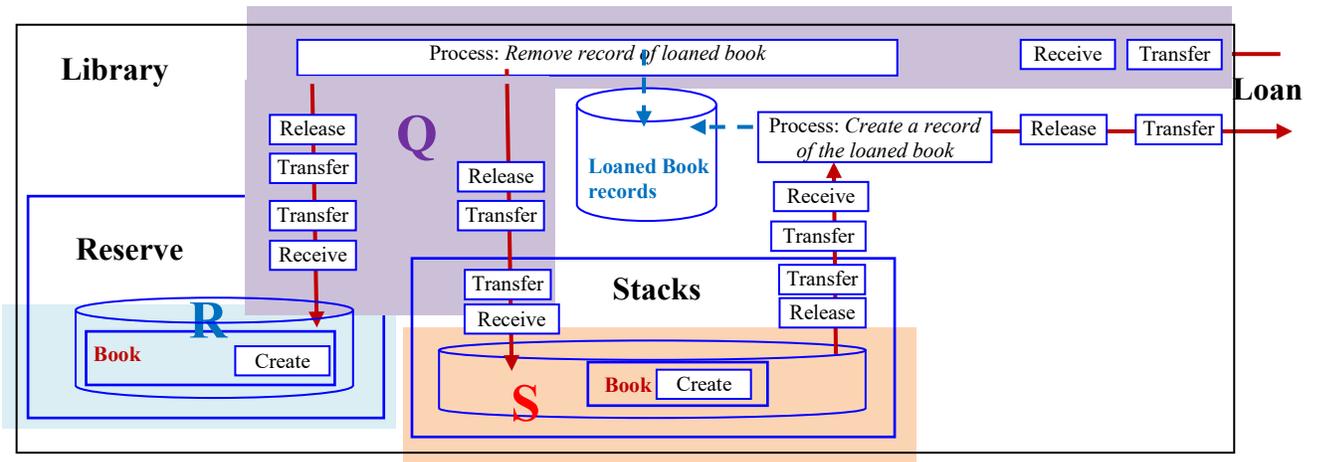

Fig. 24 The events Q, S and R. The S or R is expressed in the chronology of events diagram

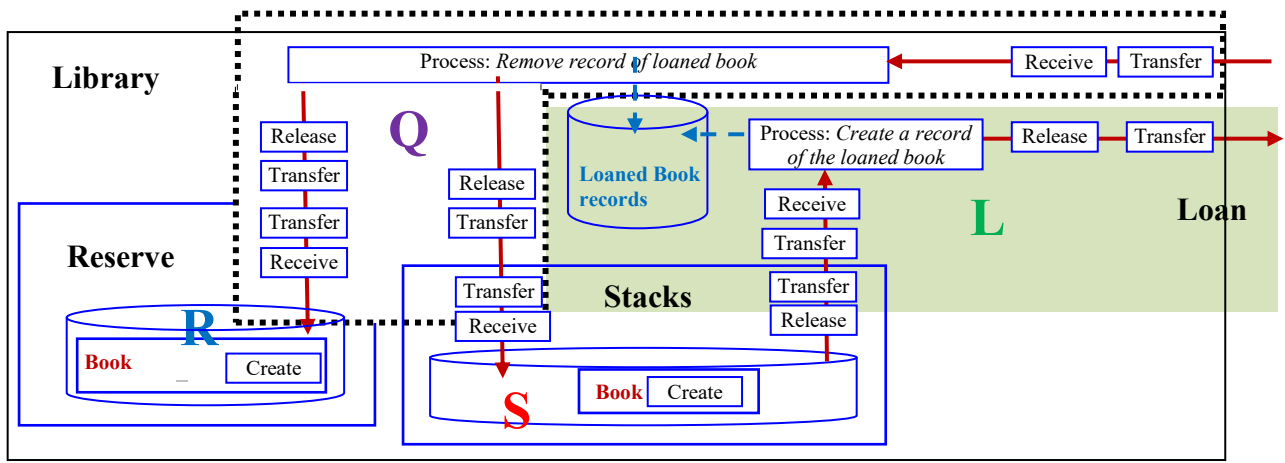

Fig. 25 The events Q (absent and L.

## VII. Conclusion

This paper explored applying propositional logic language to the high-level conceptual representation of thinging machines (TM). The aim is to build a semantic base as an intermediate interpretation of TM and propositional logic notions. Several interesting results have been developed, primarily concerning the negativity of propositions and the absence of events. The approach seems to be fruitful for TM modeling and propositional logic. This paper sets the apparatus of aligning the two fields in place, at least tentatively. No doubt, the treatment needs further scrutiny and development that will introduced in further research.